\RequirePackage{fixltx2e}
\documentclass[twocolumn,secnumarabic,superscriptaddress,amssymb, nobibnotes, aps, prb]{revtex4-1}
\usepackage[figuresright]{rotating}
\usepackage[T1]{fontenc}
\usepackage{layouts}
\usepackage{amsfonts}
\usepackage{dcolumn}
\usepackage{amssymb}
\usepackage{amsmath}
\usepackage{graphicx}
\usepackage{supertabular}
\usepackage{bm}
\usepackage{braket}
\usepackage{hyperref}
\usepackage{color,soul}
\usepackage[english]{babel}
\usepackage[inline]{enumitem}
\usepackage{multirow}
\usepackage{changes}

\let\oldAA\AA
\renewcommand{\AA}{\text{\normalfont\oldAA}}

\begin{document}

\title{Lattice dynamics effects on the magnetocrystalline anisotropy energy: application to MnBi}%

\author{Andrea Urru}
\affiliation{International School for Advanced Studies (SISSA), \\
Via Bonomea 265, 34136 Trieste (Italy).}
\author{Andrea Dal Corso}
\affiliation{International School for Advanced Studies (SISSA), \\
Via Bonomea 265, 34136 Trieste (Italy).}
\affiliation{DEMOCRITOS IOM-CNR Trieste (Italy).}

\date{\today}%

\begin{abstract}
Using a first-principles fully relativistic scheme based on ultrasoft pseudopotentials and density functional perturbation theory, we 
study the magnetocrystalline anisotropy free energy of the ferromagnetic binary compound MnBi. We find that differences in the phonon 
dispersions due to the different orientations of the magnetization (in-plane and perpendicular to the plane) give a difference between 
the vibrational free energies of the high-temperature and low-temperature phases. This vibrational contribution to the magnetocrystalline 
anisotropy energy (MAE) constant, $K_u$, is non-negligible. When the energy contribution to the MAE is calculated by the PBEsol exchange 
and correlation functional, the addition of the phonon contribution allows to get a $T = 0$ K $K_u$ and a spin-reorientation transition 
temperature in reasonable agreement with experiments. 

\end{abstract}

\maketitle
\section{Introduction}
Recently, there has been a significant effort toward the realization of rare-earth-free permanent 
magnets \cite{{REF_magnets_1},{REF_magnets_2}}. Due to its magnetic properties, such as a high Curie 
temperature, well above room temperature, and a large uniaxial magnetic anisotropy, MnBi \cite{{REF_magnets_2},
{MnBi_1},{MnBi_2},{MnBi_3}} has emerged as a promising candidate among the transition-metal-based materials. 

Below the Curie temperature, estimated to be $T_c = 680$ K \cite{{T_c},{T_c_2}}, MnBi is a ferromagnet which crystallizes 
in the NiAs structure. Its magnetocrystalline anisotropy energy (MAE) as a function of temperature is peculiar: 
at $T = 0$ K the MAE constant $K_u$ is negative, its reported experimental value being $- 0.2 $ 
MJ / m$^3$ ($\approx - 0.12 $ meV / cell) with an easy axis in the basal plane \cite{exp_K_u}, and increasing 
with $T$, unlike most magnetic systems \cite{{MnBi_2},{exp_K_u}}. At $T \approx 90$ K $K_u$ becomes positive, 
thus leading to a spin-reorientation transition: from $90$ K to $140$ K, the easy axis rotates outside 
the basal plane, and above $140$ K it is parallel to the c axis \cite{Hihara}.

Several studies, both experimental and theoretical, have been carried out during the past years to understand 
this intriguing property. Experiments studied several properties of MnBi, including the thermal 
expansion. In particular, the spin-reorientation transition comes together with a small kink in 
the lattice parameters at $T \approx 90$ K \cite{{exp_latt_1},{exp_latt_2}}, which has been interpreted as the sign of a phase transition.
Theoretical calculations, based on Density Functional Theory (DFT) within the Local Density Approximation (LDA) and the 
Generalized Gradient Approximation (GGA) for the exchange-correlation functional, correctly predict MnBi to be 
a metal and a ferromagnet in the low-temperature phase, and to have a negative $K_u$, in agreement with experiments. 
Yet, they are believed to overestimate the magnitude of $K_u$ by nearly an order of magnitude and are often not able to reproduce the 
correct behavior of $K_u$ as a function of temperature. Refs. \onlinecite{Zarkevich} and \onlinecite{Kotliar} showed that 
the treatment of correlation effects by means of the DFT+U approach is important to get the correct behavior of $K_u$ as 
a function of temperature. In particular, in Ref. \onlinecite{Kotliar} the inclusion of the thermal expansion effects on $K_u$ 
allowed to get a spin-reorientation temperature in agreement with experiments and a theoretical MAE in good agreement 
with experimental results, especially in the temperature range 150-450 K. 

More recently, in Ref. \onlinecite{Singh} it was suggested that the spin-reorientation phenomenon might be partially 
due to lattice dynamics. Such statement was supported by the calculation of the lattice dynamics contribution 
to $K_u$, obtained by averaging the MAE over configurations in which the Mn and Bi atoms were displaced 
according to the mean square atomic displacements as a function of temperature.

Recently, we extended Density Functional Perturbation Theory (DFPT) for lattice dynamics with Fully 
Relativistic (FR) Ultrasoft pseudopotentials (US-PPs) to magnetic materials \cite{DFPT_SO_mag}. The new formulation 
allows to detect differences in the phonon frequencies for different orientations of the magnetization, 
thus making possible to evaluate the vibrational free energy contribution to the MAE.

In this paper we study, by means of ab-initio techniques, the lattice dynamics of ferromagnetic MnBi for two 
different orientations of the magnetization: 
\begin{enumerate*}
\item in-plane; 
\item perpendicular to the plane. 
\end{enumerate*}
We find that the two phonon dispersions mainly differ in the high-frequency optical branches, where the 
phase with magnetic moments pointing in the out-of-plane direction shows, on average, phonon modes of 2 cm$^{-1}$ lower in 
frequency. Starting from the difference of the vibrational density of states of the two phases we compute 
the vibrational contribution to MAE. We find that, if the energy contribution to MAE is computed by the PBEsol exchange-
correlation functional, the phonon contribution is of the same order of magnitude as the ground state MAE, hence it 
plays a relevant role in the calculation of $K_u$ and to determine the spin-reorientation transition temperature $T_{SR}$. 

\section{Methods}
First-principle calculations were carried out by means of DFT \cite{{HK},{KS}} within the LDA \cite{PZ} and the 
Perdew-Burke-Ernzerhof optimized for solids (PBEsol) \cite{PBEsol} schemes for the exchange-correlation 
functional approximation, as implemented in the \texttt{Quantum ESPRESSO} \cite{{QE},{QE_2},{QE_3}} and \texttt{thermo\_pw} 
\cite{thermo_pw} packages. The atoms are described by FR US-PPs \cite{us_fr_pseudo}, with $3p$, $4s$, and $3d$ electrons 
for Mn (PPs \texttt{Mn.rel-pz-spn-rrkjus\_psl.0.3.1.UPF} and \texttt{Mn.rel-pbesol-spn-rrkjus\_psl.0.3.1.UPF }, from pslibrary 0.3.1
\cite{{pslibrary},{pslibrary_2}}) and with $6s$, $5d$, and $6p$ electrons for Bi (PPs \texttt{Bi.rel-pz-dn-rrkjus\_psl.1.0.0.UPF} and 
\texttt{Bi.rel-pbesol-dn-rrkjus\_psl.1.0.0.UPF}, from pslibrary 1.0.0 \cite{{pslibrary},{pslibrary_2}}).

MnBi crystallizes in the NiAs structure, with an hexagonal lattice described by the point group $D_{6h}$. The inclusion 
of magnetism differentiates the structures into a low-symmetry phase ($\bm{m} \perp \bm{c}$ henceforth), below $T_{SR}$, and a high-symmetry 
phase ($\bm{m} \parallel \bm{c}$ henceforth), above $T_{SR}$. In particular, the $\bm{m} \parallel \bm{c}$ phase 
is described by the magnetic point group $D_{6h} (C_{6h})$, compatible with an 
hexagonal Bravais lattice, while the $\bm{m} \perp \bm{c}$ phase has a magnetic point group $D_{2h} (C_{2h})$, compatible with 
a base-centered orthorhombic Bravais lattice. We checked the relevance of the lattice parameter $b$, which is not 
constrained by symmetry in the $\bm{m} \perp \bm{c}$ phase, and concluded that it is not crucial to make the structure more stable 
than the $\bm{m} \parallel \bm{c}$ phase, hence in the rest of the paper we use the ideal value $b = \sqrt{3} a$.
In Table \ref{t1} we summarize the data relative to the lattice constants and to the magnetic moment of Mn atoms, obtained 
with the LDA and PBEsol functionals, and compare them with previous theoretical results and with experiments. The 
LDA geometry is in good agreement with the theoretical results reported in Ref. \onlinecite{Singh}, but both lattice 
constants underestimate the experimental values: in particular, $a$ is 2 \% smaller than experiment, while $c$ is 
8 \% smaller than experiment. The PBEsol geometry gives lattice constants slightly smaller than PBE (reported in 
Ref. \onlinecite{Singh}) and experiments: $a$ and $c$ are 0.5 \% and 6 \% smaller than experiment, respectively.
The $\bm{m} \perp \bm{c}$ and $\bm{m} \parallel \bm{c}$ phases have slightly different lattice constants $a$ and $c$, but in Table \ref{t1} we report only 
one structure because the differences in the lattice constants are beyond the significative digits reported. 
In this paper we use the LDA to compute the phonon frequencies and their contribution to the MAE, while the PBEsol 
is used to compute the energy contribution to the MAE and to correct it for thermal expansion effects. The LDA and 
the PBEsol (at $T = 0$ K) calculations are performed at the geometry reported in Table \ref{t1}. The computed Mn 
magnetic moment $m_{Mn}$ is in agreement with previous calculations reported in literature \cite{{Zarkevich},{Singh}}: $m_{Mn}$ is 10 $\%$ 
(20 $\%$) smaller than experiment within the PBEsol (LDA) approximation. 

\begin{table}[t]
\centering
\begin{tabular}{c|c|c|c}
\hline 
\hline
Exchange-correlation & $a$ (\AA) & $c$ (\AA) & $m_{Mn}$ ($\mu_B$) \\
functional & & & \\
\hline
LDA (this work) & 4.16 & 5.57 & 3.2 \\
\hline 
LDA (Ref. \onlinecite{Singh}) & 4.20 & 5.54 & 3.29 \\
\hline 
GGA-PBEsol (this work) & 4.24 & 5.67 & 3.5 \\
\hline 
GGA-PBEsol (Ref. \onlinecite{Singh}) & 4.28 & 5.63 & 3.56 \\ 
\hline
GGA-PBE (Ref. \onlinecite{Singh}) & 4.35 & 5.76 & 3.69 \\
\hline 
GGA-PBE (Ref. \onlinecite{Zarkevich}) & 4.31 & 5.74 & 3.45 \\
\hline 
GGA-PBE + U (Ref. \onlinecite{Zarkevich}) & 4.39 & 6.12 & 3.96 \\
\hline 
\hline
exp. (Ref. \onlinecite{exp_latt_1}) & 4.27 & 6.05 & 3.8-4.2 \\
\hline
\hline
\end{tabular}
\caption{Computed (FR LDA and PBEsol), theoretical reference (LDA, PBEsol, PBE, and PBE+U), and experimental lattice constants and Mn magnetic moments.}
\label{t1}
\end{table}

The pseudo wave functions (charge density) have been expanded in a plane waves basis set with a kinetic energy cut-off 
of 110 (440) Ry. The Brillouin Zone (BZ) integrations have been done using a shifted uniform Monkhorst-Pack mesh \cite{k_grid} of
 $12 \times 12 \times 8$ \textbf{k}-points. The presence of a Fermi surface has been dealt with by the 
Methfessel-Paxton smearing method \cite{MP}, with a smearing parameter $\sigma = 0.015$ Ry. 
The dynamical matrices have been computed on a uniform $4 \times 4 \times 3$ \textbf{q}-points mesh, and a Fourier 
interpolation was used to obtain the complete phonon dispersions and the free energy. The latter has been obtained 
approximating the BZ integral with a  $300 \times 300 \times 300$ \textbf{q}-points mesh.  

\section{Results}
MnBi is a magnetic binary compound, in which magnetism is carried mainly by the Mn atoms, while Bi is responsible 
for a strong spin-orbit interaction. As a consequence, strong magnetocrystalline anisotropy effects are expected. 

\subsection*{Phonon dispersions}

Here we consider the phonon dispersions of MnBi with two different orientations of the magnetic moments, $\bm{m} \perp \bm{c}$ 
(in-plane) and $\bm{m} \parallel \bm{c}$ (out-of-plane), and among all the possible in-plane orientations $\bm{m} \perp \bm{c}$, 
we choose $\bm{m} \parallel \bm{a}$, $\bm{a}$ and $\bm{c}$ being the primitive vectors of the hexagonal Bravais lattice). 
The presence of a magnetization leads to a difference in the Bravais lattice the magnetic point group is compatible with, 
as discussed in the previous Section. In order to compare the phonon dispersions in the same BZ, we choose to set the 
geometry in the base-centered orthorhombic Bravais lattice, which is compatible with the low-symmetry phase 
(magnetic point group $D_{2h} (C_{2h})$). 

\begin{figure}
\centering
\includegraphics[width=0.5\textwidth]{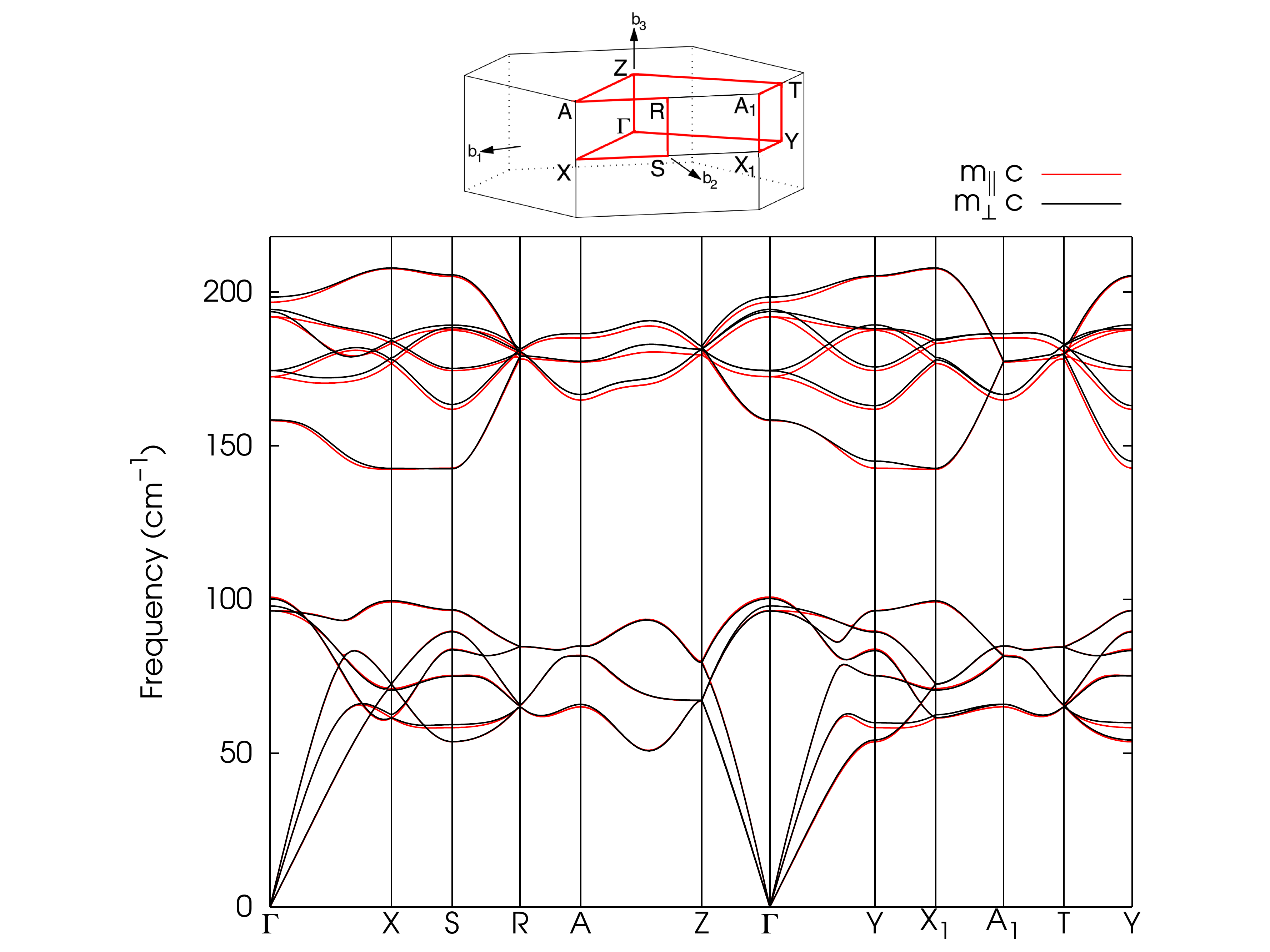}
\caption{Computed FR LDA phonon dispersions of MnBi with magnetic moments oriented in plane ($\bm{m} \perp \bm{c}$, 
black line), and out of plane ($\bm{m} \parallel \bm{c}$, red line).}
\label{f1}
\end{figure}

The phonon dispersions are illustrated in Fig. \ref{f1}. The phonon modes are split in two groups, separated 
by a gap. The low-frequency branches (up to $\sim 100$ cm $^{-1}$) are dominated by displacements of the heavy 
element Bi, while the high-frequency branches (from $\sim 150$ cm $^{-1}$ to $\sim 200$ cm $^{-1}$) are mainly 
displacements of the Mn atoms. The main difference between the phonon frequencies of the two phases is a rigid
shift: the phonon frequencies of the phase with in-plane magnetization are higher than those of the phase with 
out-of-plane magnetization. The shift is about $0.5$ cm $^{-1}$ in the low-frequency branches, while it is about 
2 cm $^{-1}$ in the high-frequency branches. Moreover, there are differences due to symmetry. The system with 
in-plane magnetization has lower symmetry and some modes, degenerate when the magnetization is along $\bm{c}$,
split. As an example, in Table \ref{t3} we report the phonon modes at Z and X$_1$. 
At Z, in the phase with $\bm{m} \parallel \bm{c}$ there are two groups of four-fold degenerate modes, which 
become four couples of degenerate modes in the phase $\bm{m} \perp \bm{c}$: the splittings are quite small, in 
the range 0.02-0.08 cm$^{-1}$. At X$_1$, in the configuration $\bm{m} \parallel \bm{c}$  there are four two-fold 
degenerate modes, which split from 0.1 cm$^{-1}$ to 1 cm$^{-1}$. Similar splittings are found also along the other 
high-symmetry lines.

\begin{table}[t]
\centering
\begin{tabular}{c|cc|cc}
  \hline 
  \hline
  q & \multicolumn{2}{c}{$\bm{m} \parallel \bm{c}$} & \multicolumn{2}{c}{$\bm{m} \perp \bm{c}$} \\
  \hline
  \multirow{5}{*}{Z} & $\nu$ (cm $^{-1}$) & degeneracy &  $\nu$ (cm $^{-1}$) & degeneracy \\
  \cline{2-5}
  & \multirow{2}{*}{67.135} & \multirow{2}{*}{4} & 67.212 & 2 \\
  & & & 67.231 & 2 \\
  \cline{2-5}
  & \multirow{2}{*}{180.245} & \multirow{2}{*}{4} & 181.263 & 2 \\
  & & & 181.340 & 2 \\
  \hline
  \multirow{8}{*}{X$_1$}
  & \multirow{2}{*}{61.427} & \multirow{2}{*}{2} & 61.554 & 1 \\
  & & & 62.503 & 1 \\
  \cline{2-5}
  & \multirow{2}{*}{72.461} & \multirow{2}{*}{2} & 72.467 & 1 \\
  & & & 72.577 & 1 \\
  \cline{2-5}
  & \multirow{2}{*}{176.735} & \multirow{2}{*}{2} & 178.156 & 1 \\
  & & & 178.556 & 1 \\
  \cline{2-5}
  & \multirow{2}{*}{184.374} & \multirow{2}{*}{2} & 185.195 & 1 \\
  & & & 185.395 & 1 \\
  \hline
  \hline
\end{tabular}
\caption{Computed FR LDA phonon frequencies at high-symmetry points Z and X$_1$ for the configurations $\bm{m} \parallel \bm{c}$ 
and $\bm{m} \perp \bm{c}$. Only the degenerate modes are shown for the configuration $\bm{m} \parallel \bm{c}$, and how the 
degeneracy is lowered or lifted if $\bm{m} \perp \bm{c}$.}
\label{t3}
\end{table}

\subsection*{MAE}

In previous works, the spin-reorientation transition, due to the change of $K_u$ from negative to positive at $T_{SR} \approx 90$ K, 
has been explained as an effect of thermal expansion of the crystal parameters $a$ and $c$. 

\begin{figure}[t]
\centering
\includegraphics[width=0.5\textwidth]{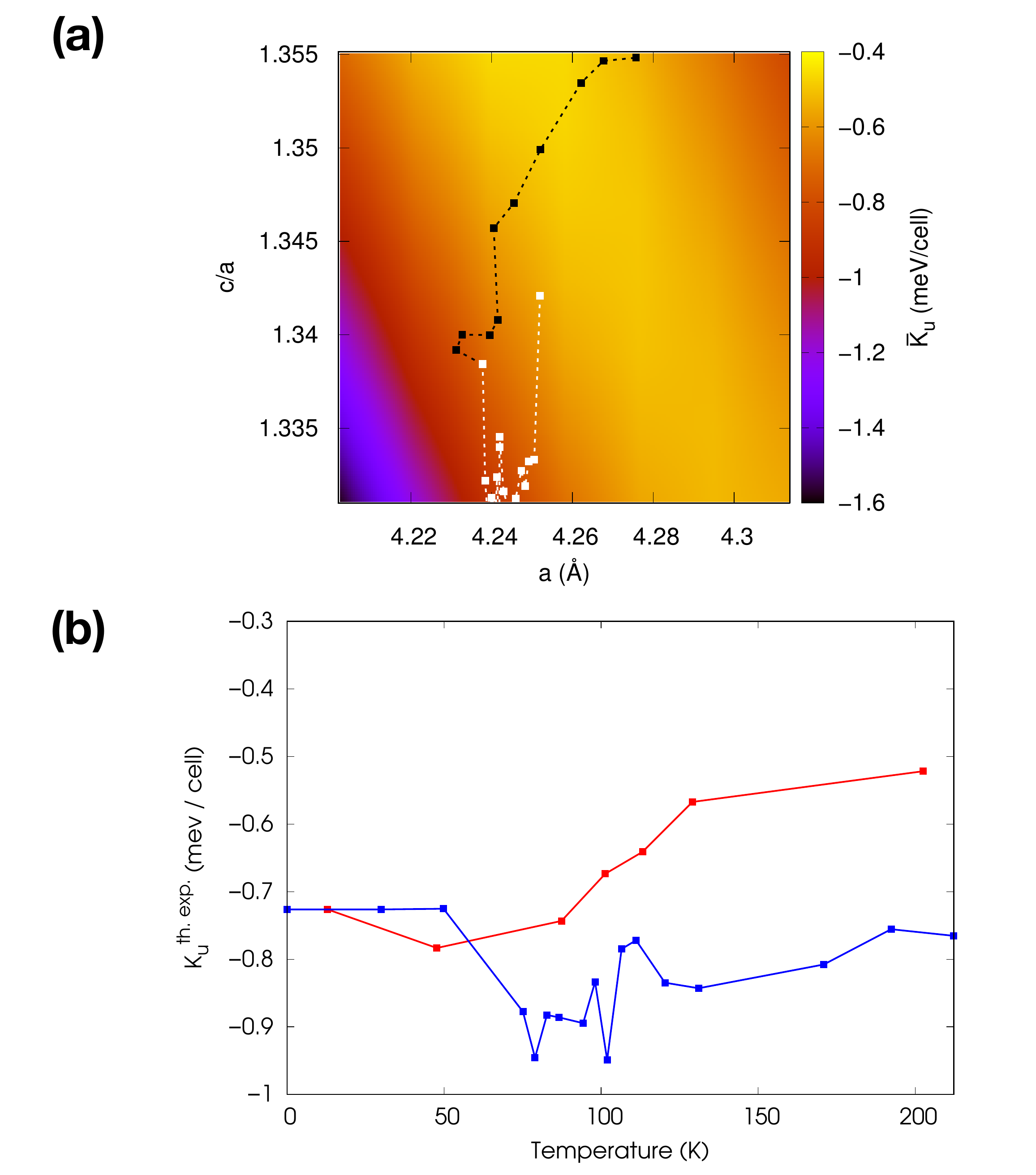}
\caption{Energy contribution to the MAE constant, $\bar{K_u}$, computed in the FR PBEsol scheme. (a) $\bar{K_u}$ as a function of lattice constants. 
Black and white dashed lines represent the geometries at different temperatures, obtained from Refs. \onlinecite{exp_latt_1} and 
\onlinecite{exp_latt_2}, respectively, as explained in the main text. (b) $K_u^{th. exp.}$ as a function of temperature. Red 
and blue lines represent the values of $K_u^{th. exp.}$ obtained from the black and white dashed lines of Fig. 2 (a), respectively.}
\label{f2}
\end{figure}

In Refs. \onlinecite{Zarkevich} and \onlinecite{Kotliar} $K_u^{th. \, exp.}$, the function $K_u$ obtained accounting for thermal 
expansion effects, has been computed within the LSDA $+$ SO (spin-orbit)$+$ U scheme  using the experimental lattice constants as a function 
of the temperature, reported in Refs. \onlinecite{{exp_latt_1},{exp_latt_2}} and finding in this way a good agreement with experiments.
Here instead we compute $K_u^{th. \, exp.}$ within the FR PBEsol scheme. In Fig. \ref{f2}(a) we show $\bar{K}_u$, the contribution to 
$K_u$ given by the energy difference of the electronic groundstates,
for a mesh of geometries and on top of it we indicate with black and white points the thermal expansion data added to the theoretical 
PBEsol $T = 0$ K crystal parameters. The resulting $K_u^{th. \, exp.}$ is reported in Fig. \ref{f2}(b). $K_u^{th. \, exp.}$ is negative 
and slightly increases with increasing $a$ and $c$. Yet, such an increase is not sufficient to cross the value $K_u^{th. exp.} = 0$ 
because the $T = 0$ K  energy difference ($\approx -0.73$ meV/cell) is significantly lower than the experimental value ($\approx -0.15$ meV 
/ cell), similarly to what found within the LDA and GGA approximations in Refs. \onlinecite{Zarkevich} and \onlinecite{Kotliar}, and 
because the energy difference landscape shows a local maximum of about $-0.4$ meV / cell. 

In addition to the thermal expansion effect we consider also the effect of the lattice dynamics on $K_u$: in fact, since the 
two phases have slightly different phonon frequencies, they have different vibrational entropies that give a temperature-dependent 
contribution to the MAE defined as the difference of the vibrational free energies of the two phases. We write the constant $K_u$ as \cite{vib_mag_separate}: 
\begin{equation}
K_u = K_u^{th. \, exp.} + K_u^{vib} + K_u^{mag}, 
\label{eq1}
\end{equation} 
where $K_u^{th. \, exp.}$ is the thermal expansion contribution, $K_u^{vib}$ is the lattice dynamics contribution, and $K_u^{mag}$ is the 
magnon contribution, which we do not consider in the present work. The term $K_u^{vib}$ can be computed from the phonon frequencies using 
the harmonic approximation: 
\begin{align}
K_u^{vib} & = \int_0^{+ \infty} d \omega \, \frac{1}{2} \, \hbar \omega \, \left[ g_{\perp}(\omega) - g_{\parallel} (\omega) \right] \nonumber \\
& + k_B T \, \int_0^{+ \infty} d \omega \, \left[ g_{\perp}(\omega) - g_{\parallel} (\omega) \right] \ln \left( 1 - e^{- \hbar \omega / k_B T} \right),
\label{eq2}
\end{align}
where $ g_{\perp}(\omega)$ ($ g_{\parallel}(\omega)$) is the phonon density of states relative to the phase with $\bm{m} \perp \bm{c}$ ($\bm{m} \parallel \bm{c}$). 
\begin{figure}[t!]
\centering
\includegraphics[width=0.5\textwidth]{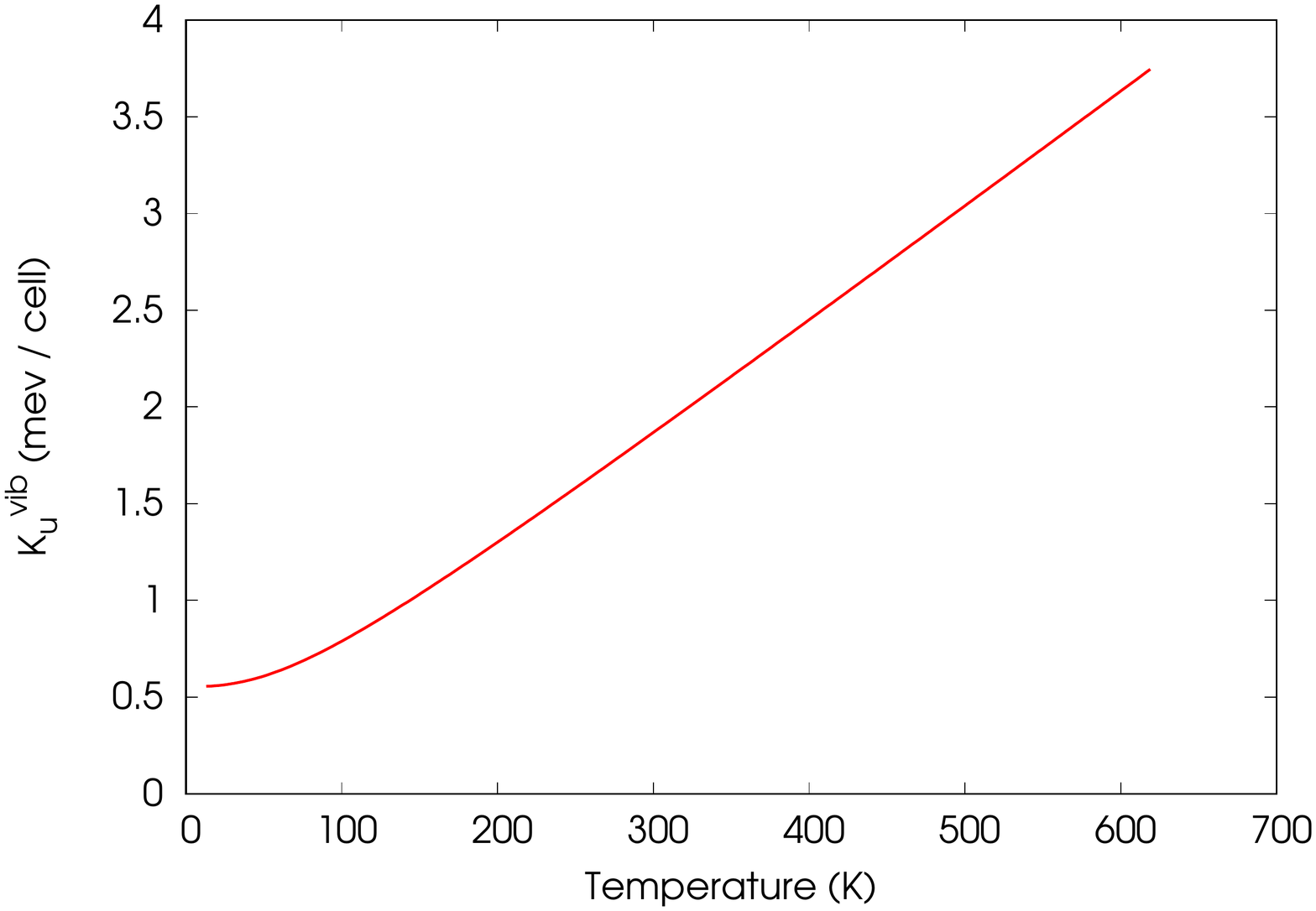}
\caption{Vibrational contribution to the MAE constant $K_u$, computed from the phonon frequencies within 
the harmonic approximation (Eq. \eqref{eq2})}
\label{f3}
\end{figure}
\begin{figure}[t!]
\centering
\includegraphics[width=0.5\textwidth]{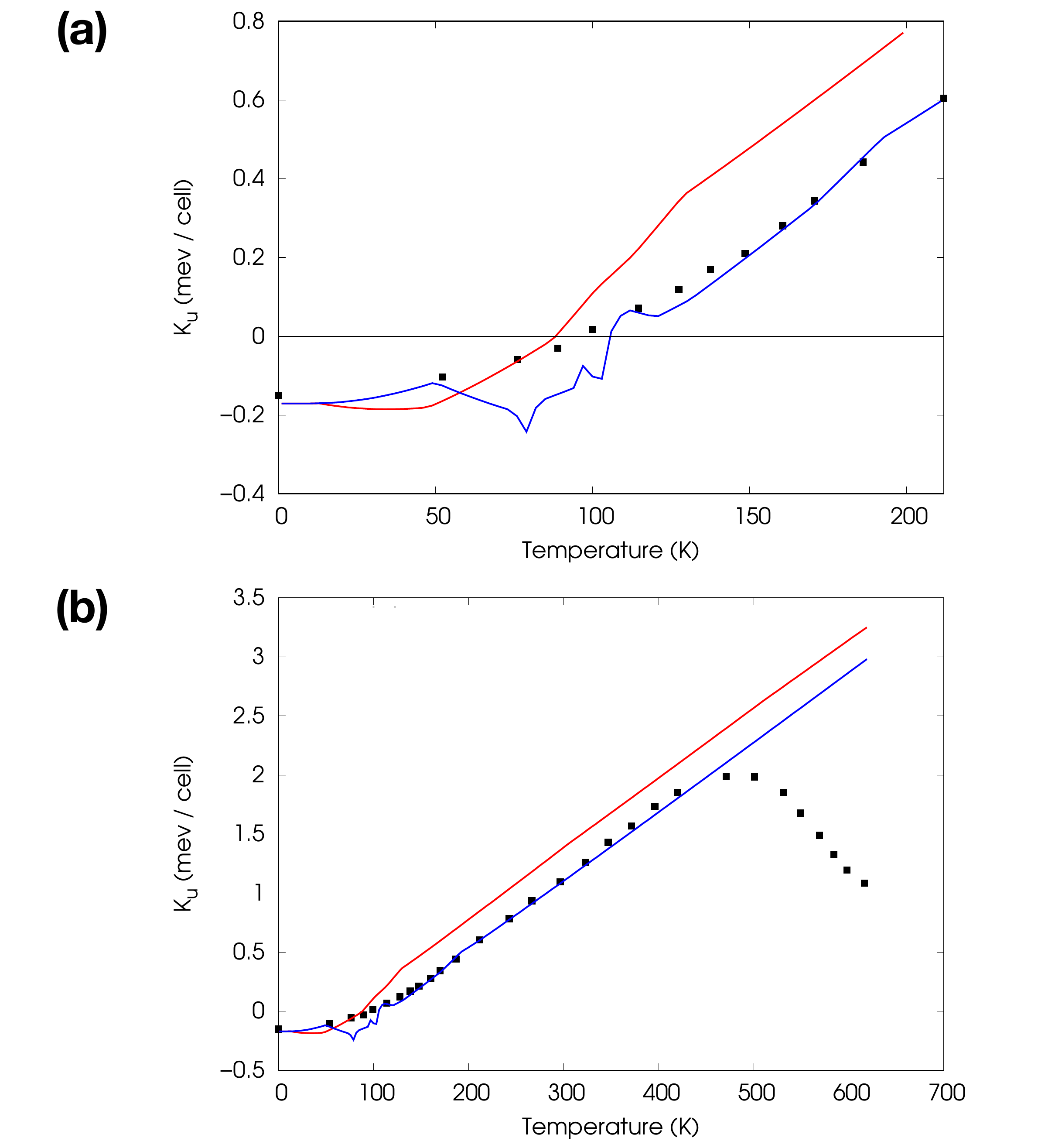}
\caption{Comparison between computed and experimental MAE constant $K_u$. Black squares represent experimental data \cite{exp_K_u}, red and 
blue lines represent the total $K_u$ computed, where the $K_u^{th. exp.}$ contribution has been obtained from Refs. \onlinecite{exp_latt_1} and 
\onlinecite{exp_latt_2}, respectively, as explained in the main text. (a) Detailed comparison in the temperature range $0-200$ 
K, to highlight $T_{SR}$. (b) Comparison in a wider range of temperatures ($0-600$ K).}
\label{f4}
\end{figure}
In our case $K_u^{vib}$ is always positive and increases with increasing temperature, as shown in Fig. \ref{f3}: its magnitude is comparable 
with that of $K_u^{th. \, exp.}$, hence it gives a crucial contribution in determining the MAE constant $K_u$. 

In Figs. \ref{f4}(a-b) we show the total $K_u$ defined in Eq. \eqref{eq1} and compare it with the experimental data. By comparison with Fig. 
\ref{f2}(b), it is clear that the addition of the vibrational contribution is important. In particular, the zero-point vibrational free 
energy difference allows to have a $T = 0$ K value of $K_u$ in good agreement with experiments and, together with the thermal contribution 
(second term of Eq. \eqref{eq2}), to get a transition temperature $T_{SR}$ in reasonable agreement with the experiments ($T_{SR} \approx 
90$ K and $T_{SR} \approx 110$ K with the data given by Refs. \onlinecite{exp_latt_1} and \onlinecite{exp_latt_2}, respectively). Moreover, 
the vibrational contribution allows to have a fair agreement with experimental data \cite{exp_K_u} in a rather large temperature range, 
as shown in Fig. \ref{f4}(b). At variance with Ref. \onlinecite{Kotliar}, we do not find a maximum in $K_u$ because the 
$T=0$ K geometry corresponds to the theoretical geometry and because the vibrational contribution increases with $T$. At high temperatures 
($T > 500$ K) our results do not agree with the experiment, suggesting that additional contributions may become important in this temperature 
range, in particular the term $K_u^{mag}$ in Eq. \eqref{eq1}, which can result in a decreasing $K_u$ with $T$ \cite{magnons}. Moreover, 
we mention that in the high-temperature limit additional effects not included in Eq. \eqref{eq1}, as the magnon-phonon coupling 
\cite{{mag-phon_1},{mag-phon_2}}, might be non-negligible.

\section{Conclusions} By means of recent developments in first-principles theoretical tools, we have studied the lattice dynamics of the MnBi 
ferromagnet for two orientations of the magnetization, $\bm{m} \perp \bm{c}$ and $\bm{m} \parallel \bm{c}$. We have shown 
that the differences in the phonon frequencies give rise to a contribution to the MAE that is 
comparable with the electronic one. We have found that the vibrational contribution is relevant to explain the behavior of the 
MAE constant $K_u$ as a function of temperature. We could also get an estimate of the spin-reorientation temperature $T_{SR}$ in fair agreement 
with the experimental value using the PBEsol approximation to evaluate the energy contribution to the MAE. The use of other functionals such as LDA would 
give instead a lower value of the MAE at $T = 0$ K than that predicted by PBEsol, and would lead to predict $T_{SR} \approx 500$ K. The use of the Hubbard U parameter would result, instead, in a higher value of $K_u$ at $T = 0$ K \cite{Kotliar} and, as a consequence, 
we would get $K_u > 0$ when adding the vibrational contribution, thus the spin-reorientation transition would not be explained.

In Ref. \onlinecite{Singh} a first estimate of the vibrational contribution to MAE was given, which led to $T_{SR} \approx 450$ K, far from the 
experimental value. In this work we have found a vibrational contribution of the same order of magnitude as found in Ref. \onlinecite{Singh}, 
and we have included also the zero-point vibrational free energy difference, which gives an important contribution and leads to an estimated 
spin-reorientation temperature $T_{SR} \approx 100$ K, when used together with the energy MAE given by PBEsol. 

\section*{Acknowledgments}

Computational facilities have been provided by SISSA through its Linux Cluster and ITCS and by CINECA through 
the SISSA-CINECA 2019-2020 Agreement.

\end{document}